**Hybrid artificial intelligence echogenic components-based diagnosis of adnexal masses on ultrasound**

Roni Yoeli-Bik, M.D.[1a], Heather M. Whitney, Ph.D.[2a*], Hui Li, Ph.D.[2], Agnes Bilecz, Ph.D.[1], Jacques S. Abramowicz, M.D.[1], Li Lan, M.S.[2], Ryan E. Longman, M.D.[1], Maryellen L. Giger, Ph.D.[2b], and Ernst Lengyel, M.D., Ph.D.[1b]

[1] Departments of Obstetrics and Gynecology and [2] Radiology

[a] co-first authors
[b] co-senior authors

* Corresponding author
Heather M. Whitney, Ph.D., hwhitney@uchicago.edu
The University of Chicago, 5841 S. Maryland Avenue, Chicago, IL, USA 60637
tel: 773-834-7769

## Abstract

**Background:** Adnexal masses are heterogeneous and have varied sonographic presentations, making them difficult to diagnose correctly.

**Purpose:** Our study aimed to develop an innovative hybrid artificial intelligence/computer aided diagnosis (AI/CADx)-based pipeline to distinguish between benign and malignant adnexal masses on ultrasound imaging based upon automatic segmentation and echogenic-based classification.

**Methods:** The retrospective study was conducted on a consecutive dataset of patients with an adnexal mass. There was one image per mass. Mass borders were segmented from the background *via* a supervised U-net algorithm. Masses were spatially subdivided automatically into their hypo- and hyper-echogenic components by a physics-driven unsupervised clustering algorithm. The dataset was separated by patient into a training/validation set (95 masses; 70%) and an independent held-out test set (41 masses; 30%). Eight component-based radiomic features plus a binary measure of the presence or absence of solid components were used to train a linear discriminant analysis classifier to distinguish between malignant and benign masses. Classification performance was evaluated using the area under the receiver operating characteristic curve (AUC), along with sensitivity, specificity, negative predictive value, positive predictive value, and accuracy at target 95% sensitivity.

**Results:**
The cohort included 133 patients with 136 adnexal masses. In distinguishing between malignant and benign masses, the pipeline achieved an AUC of 0.90 [0.84, 0.95] on the training/validation set and 0.93 [0.83, 0.98] on the independent test set. Strong diagnostic performance was observed at the target 95% sensitivity.

**Conclusions:** A novel hybrid AI/CADx echogenic components-based ultrasound imaging pipeline can distinguish between malignant and benign adnexal masses with strong diagnostic performance.

## Introduction

Ovarian cancer is the most lethal gynecological malignancy and the fifth-leading cause of cancer deaths among women,[1] resulting in substantial interest in improving noninvasive diagnosis. Ultrasound imaging is the initial modality for diagnosing adnexal masses because it is safe, widely available, quick to complete, and relatively cheap.[2] However, accurate diagnosis of adnexal masses presents significant challenges due to the low incidence of ovarian cancer and the frequent occurrence of ovarian masses. Indeed, the lifetime risk of being diagnosed with an adnexal mass is up to 35% in premenopausal and 17% in postmenopausal patients.[3] Adnexal masses are heterogeneous, and benign, borderline, and malignant masses often share similar morphologic characteristics.[4,5] A large meta-analysis[6] reported a pooled sensitivity and specificity of 0.93 and 0.89, respectively for the subjective classification of adnexal masses on ultrasound imaging. Some qualitative ultrasound-based risk models have been developed that have high sensitivity, but they have only moderate specificity.[6–8] Therefore, surgery, which frequently results in benign findings[9] and carries risks of patient morbidity, may be required for a definitive diagnosis. Currently, sonographically indeterminate masses may be further evaluated by magnetic resonance imaging (MRI) examinations, which provide enhanced soft tissue characterization and can potentially reduce false positives and unnecessary surgeries for asymptomatic patients with benign masses.[2,10,11] However, while MRI imaging can have a high negative predictive value and relatively high specificity,[3,12–14] its sensitivity is still limited, and this imaging modality is expensive, time-consuming, and not widely available. Consequently, improving the performance of sonographic assessments is of clinical interest, and with recent developments in computational power and data collection, more quantitative imaging assessments that may lead to improved noninvasive diagnostic accuracy are now possible.[15,16]

Artificial intelligence/computer-aided diagnosis (AI/CADx) models are based on quantitative image-based phenotypes (i.e., radiomic features) correlated with malignant and benign phenotypes and can potentially improve imaging diagnostic accuracy.[17] Due to the quantitative nature and use of algorithms, AI/CADx models may decrease interpretative variability and human error and allow for better treatment planning.[15] However, AI/CADx-based diagnostic models are not part of standard clinical practice for adnexal mass diagnosis. Given the heterogeneous characteristics of adnexal masses, even in the same histopathologic subtype, extracting quantitative imaging features that reflect and correlate with intra-mass heterogeneity and the underlying biology and tissue architecture may provide robust automatic adnexal mass assessments.

The purpose of this study was to develop an AI/CADx-based method to distinguish between benign and malignant adnexal masses on grayscale ultrasound imaging. Our approach includes a combination of AI methodologies (supervised deep learning and unsupervised machine learning) developed from a collaboration between researchers in medical physics and in gynecologic oncology. The proposed pipeline, which was developed with clinical explainability and ease of use as a priority, includes radiomic features extracted from echogenic intra-mass components to assess adnexal masses using AI/CADx.

**Materials and Methods**

***Study Patients and Mass Characteristics***

This retrospective single-center study was conducted at The University of Chicago. The study cohort was retrieved from a previously described clinicopathologic database that included more than 500 consecutive patients with adnexal masses and available ultrasound imaging (2017-2022).[18] The database has since been updated with 8 additional consecutive months of patient collection (11/2022-06/2023) under an approved HIPAA-compliant protocol by the institutional review board. Exclusion criteria by patient, mass, and imaging levels were followed for the AI development pipeline[19] (**Figure 1**). Only patients with surgical evaluations, histopathologic findings of adnexal mass origin, and ultrasound imaging conducted at the institution were included. We required one high-quality, representative transvaginal grayscale image per mass and excluded images with undefined mass borders and images with measurement sonographic markups. Patients with bilateral masses, that were malignant and met the study inclusion criteria, were included in the AI dataset, resulting in 3 additional masses. Borderline ovarian tumors were grouped with malignant masses for diagnostic performance and statistical analyses because borderline tumors also require surgery.

***Image Acquisition***

Sonographic images for the evaluation of adnexal masses had been clinically acquired using ultrasound machines (GE Voluson E8 or E10 or Samsung Elite WS80). The images were retrospectively retrieved in DICOM format and fully de-identified for the current study.

***Mass Segmentation and Feature Extraction***

The previously developed physics-driven segmentation model[19] applies a user-provided bounding box around the region of interest (adnexal mass) as input to (1) segment the masses from the image background using a supervised deep learning (DL) U-net model and (2) separate them into intra-mass relative hypo- and hyperechogenic components using an unsupervised machine learning fuzzy c-means algorithm (**Figure 2**). Eight component-based human-engineered radiomic features were extracted from each mass (**Table 2**). The features describe component morphology (spatial variations in pixel values near the edge of each component), geometry (shape and size), and texture (the spatial relationships between image pixels in terms of the change in intensity patterns and gray levels) (**Figure 2**).[20–23] These features had been identified separately from this study for their ability to individually distinguish between malignant and benign masses on ultrasound imaging, i.e., algorithmic feature selection was not conducted in this study. Four features based on mass morphology were calculated to describe the hypoechogenic components (one feature) or the hyperechogenic components (three features). Two features were based upon the geometry of the components: one feature was measured as the fraction of pixels within the mass that was within the hypoechogenic component, while one feature (which we term a 'proportion' feature) was calculated as (Eq. 1)

$$\frac{D_{eff,hypo}}{D_{eff,hyper}} \frac{A_{hyper}}{A_{hypo}}$$
Eq. 1

where $D_{eff}$ is the effective diameter of each component (i.e., the diameter of a circle with the same area as the component), $A$ refers to the area (in pixels) of each component, and the subscripts *hypo* and *hyper* refer to the relative hypo- and hyper-echogenic components, respectively. Two features, based upon the relative textures of difference entropy and correlation, were calculated as (Eq. 2)

$$\frac{f_{hypo}}{f_{hyper}}$$
Eq. 2

where $f$ is the radiomic feature from each component. A binary feature (i.e., present or absent) for the presence of solid components or lack thereof was determined from an expert manual review of each case. Thus, a total of 9 features were used in the study.

***Mass Classification***

The dataset was manually split by patient into a classification training/validation set (95 masses; 70%) and an independent held-out test set (41 masses; 30%), to match stratified adnexal pathologies and clinical parameters (menopausal status and race) between the two sets. Using only the training set data, a linear discriminant analysis classifier was trained using the nine extracted features to yield a likelihood of malignancy. This classifier was applied to the independent held-out test set. Pathology records served as the ground truth reference standard for classification. Feature importance was evaluated using Shapley analysis[24] (bee swarm and bar charts). For the task of classification of masses as malignant or benign, the figure of merit was the area under the receiver operating characteristic (ROC) curve (AUC),[25] calculated using the proper binormal model.[26] We also calculated the empirical ROC curve. Diagnostic performance was also evaluated at target 95% sensitivity, yielding corresponding specificity, positive predictive value (PPV), negative predictive value (NPV), and accuracy.

The likelihood of malignancy of individual masses, along with internal segmentation performance, was reviewed by clinicians (JSA, RYB, and REL) for qualitative correspondence with ultrasound images as well as gross and histological pathologies. This was done because of the heterogeneous nature of the masses within each mass type and the importance of understanding the biophysical basis of AI/ML pipelines in medical imaging, including radiomics.

### *Statistical Analysis*

The median and 95% confidence interval (CI) of the AUC, sensitivity, specificity, PPV, NPV, and accuracy were determined for the training/validation set and for the separate test set by *a posteriori* bootstrapping of the classifier output 2000 times, i.e., randomly sampling with replacement. Statistical analyses were performed in MATLAB (MATLAB 2022b, MathWorks).

### **Results**

### *Patient and Mass Characteristics*

The research database had 594 patients with adnexal masses.[18] The inclusion and exclusion criteria at the patient, adnexal mass, and image levels (**Figure 1**) resulted in a final dataset of 133 unique patients with 136 adnexal masses (**Table 1 and Table 2**). Three patients had bilateral malignant masses. The malignant tumor prevalence was 27.9% (38/136). The mean (range) patient age was 45 (20-82) years, and 39.8% (53/133) of the patients were postmenopausal.

### *Mass Classification Performance*

The radiomic features used for classifying adnexal masses are shown in **Table 3**, reflecting morphology, shape, size, and texture qualities. The Shapley bee swarm chart in its traditional color coding by feature value (**Figure 3, top**) indicates the trend of feature value with Shapley value (a measure of feature importance). For all but one radiomic feature, the feature values were generally inversely related to Shapley value: a high feature value was associated with a low Shapley value (i.e., more predictive of benign masses), and vice versa. The opposite was true for margin sharpness variance from the hyperechoic component, where a high feature value was associated with a high Shapley value (i.e., more predictive of malignant masses), and vice versa. When the same Shapley bee swarm chart is viewed with color coding by benign or malignant pathology (**Figure 3, bottom**), the substantial differences in feature importance among benign and malignant masses and within the same group are more evident, supporting the integration and analysis of multiple components-based features to improve classification performance. We investigated these differences further by examining the feature importance for four example masses (**Figure 4**), two malignant and two benign. For the first example malignant mass (labeled M1 for this study), the margin sharpness variance of the hyperechoic component is almost important as the binary solid feature. For the first example benign mass (labeled B1 for this study), the proportion feature of effective diameter was the most important feature, while it was not as important for the other masses in this example set. Example malignant mass M2 and example benign mass B2 have both similar feature values (indicated by the color of the bars) and similar importance (indicated by the length of the bars) of two features, (a) the ratio of the difference entropy between echogenic components and (b) the proportion of effective diameter. However, the ratio of the fraction of pixel values in B2 is more important for its classification than for M2. Altogether, the Shapley values from these example masses indicate the complex nature of classification for heterogeneous adnexal masses, both between mass types and within each type, supporting the development of an integrated model that is based on the different echogenic mass components analysis.

The corresponding ROC curves for independent test set, as well as the training/validation set, are shown in **Figure 5** for both empirical (raw) and fitted curves. The AUC in distinguishing between malignant and benign masses was 0.90 [0.84, 0.95] in the training/validation set and 0.93 [0.83, 0.98] in the independent test set using the proper binormal model. The AI/CADx model achieved the target 95% sensitivity, along with a very high NPV of 0.997 [0.935, 1.000] and relatively high specificity of 0.71 [0.53, 0.84] and PPV of 0.58 [0.47, 0.71] in the independent test set (**Table 4**).

The review of the camera images and pathology of individual masses showed that the automatic intra-mass segmentations and some of the radiomic features corresponded with mass characteristics observable in qualitative image assessment, gross pathology, and histopathological examination (**Figure 6**). Furthermore, the radiomic features captured unique characteristics that were useful for diagnosis but not easily viewable in qualitative image assessment. These are apparent in the examples pictured in **Figure 6** and described below, which portray the use of corresponding qualitative and quantitative analyses.

Benign serous cystadenofibromas, which are often difficult to diagnose,[4] usually appear on greyscale and Doppler ultrasound imaging as cystic masses that have avascular solid papillary projections with posterior acoustic shadowing.[27] These qualitative imaging characteristics correspond with the known gross pathology since cystadenofibromas contain cysts larger than 1 cm and variable amounts of solid areas, often with simple and broad papillae. On histology, cysts lined by a single layer of serous epithelium are surrounded by dense fibromatous stroma.[28] These characteristics were also recognized by the automatic segmentation, as seen in **Figure 6 A-B**. Some of the radiomic features used in this study (measuring size and shape) reflect geometric characteristics of hyperechogenic components, including papillary projections such as in these benign serous cystadenofibromas. Other radiomic features, particularly the edges of the hyperechogenic components as measured by morphology features, correspond to characteristics of the solid papillary projections for this type of benign mass. These features characterize three different qualities of the hyperechogenic components (margin sharpness variance, margin sharpness mean, and variance of the radial gradient histogram), broadening and quantifying the characteristics of distinguishing them from solid papillary projections in malignant masses. Furthermore, the ratio of texture features between components captures distinctive aspects of these masses that are not readily viewed through visual inspection of ultrasound images.

In comparison, for malignancies analysis different radiomics features were informative. For example, patients with advanced-stage high-grade serous ovarian cancers (HGSOC), the most common ovarian cancer subtype, often present with bilateral irregular solid or multilocular-solid masses (**Figure 6 C**), ascites, and upper abdominal disease.[29] High vascular flow and areas of necrosis in the solid elements are common on Doppler ultrasound imaging.[29] Another rare type of ovarian cancer, malignant Sertoli-Leydig cell tumors, are of sex-cord stromal origin. On ultrasound imaging, they may have variable appearances; they are usually purely solid or multilocular-solid masses with areas of packed small cystic locules in the solid elements,[30] as seen in **Figure 6 D**. High vascular flow on Doppler imaging is often present. The size and shape of radiomic features used in this study reflect the larger fraction of hyperechogenic components often seen in malignant adnexal masses. The use of morphology features from these hyperechogenic components and the ratio of textures between the hypo- and hyperechoic components for distinguishing malignant from benign masses additionally emphasizes that the edges of the hyperechoic components and the relative textures of the components characterize the malignant masses in aspects not easily seen in a qualitative review of the images and help distinguish them from benign masses.

Overall, the use of radiomic features demonstrates that quantitative imaging characteristics, which are not readily apparent to the eye (i.e., edges of components, the relative nature of features between components, and the merging of the features by a classifier), provide measures of adnexal masses that are unique and supplement the existing qualitative sonographic review frameworks. Additional examples of sonographic and AI-based component analyses from the training/validation set are presented in **Supplementary Figure 1.**

## Discussion

Adnexal masses are common in both pre and postmenopausal patients. Ultrasound is the preferred initial imaging modality for characterizing these masses, but image interpretation can be difficult, and masses are frequently classified as indeterminate. In this study, an AI/CADx-based pipeline to differentiate between malignant and benign adnexal masses on ultrasound images, based upon characteristics of edges, geometry, and characteristics of hypo- and hyperechogenic components, demonstrated overall strong classification performance with an overall AUC of 0.93 in the independent test set (**Figure 5**). At a target 95% sensitivity to maximize cancer diagnosis, the specificity was 0.71 [0.53,0.84] (**Table 4**), which may suggest that further pipeline optimization is warranted to minimize false positive results. However, at the same target, the NPV was 0.997 [0.935,1.000], clinically reassuring that a negative test is accurate and that no cancer is misclassified as a benign mass. In routine clinical practice, MRI is often used as a secondary imaging modality when the ultrasound assessment is suboptimal or indeterminate (O-RADS ultrasound scores of 3 or 4). In a large prospective multicenter cohort of patients with sonographically indeterminate adnexal masses, diagnostic pelvic MRI had an NPV of 98% (at 18% malignancy prevalence).[10] Our study presents a low-complexity model with a high and reassuring NPV based on ultrasound imaging. This imaging is widely available, much cheaper than MRI, and does not require additional interpretative skills.

Our findings are consistent with previously published work by others on AI-based classification of adnexal masses using different imaging modalities.[31,32] A recent meta-analysis of AI-based systems found a pooled sensitivity of 0.91, a pooled specificity of 0.87, and an AUC of 0.95 with ultrasound.[31] However, a more comprehensive pipeline, including mass segmentation and classification[33,34] and assessments by different mass components[34–36] on ultrasound imaging, was seldom explored. Moreover, none of these studies analyzed the heterogeneous nature of adnexal masses by the relative echogenicity of mass components under the same unsupervised pipeline. The performance of our AI/CADx-based model suggests that a comprehensive pipeline using human-engineered component-based feature extraction and analysis, while leveraging supervised DL and unsupervised ML tools for automatic mass segmentation, can provide a comprehensive methodology for sonographic adnexal mass assessment independent of human variability and more accurate mass diagnosis.

The features used in this study inform our understanding of radiomic properties useful in classifying adnexal masses. Variations of these features were previously used to diagnose whole breast lesions on ultrasound images.[20–23] In the current study, the edges of the intra-mass components, characterized through morphology features, were important for classification. This may indicate that the boundaries of intra-mass components portray characteristics that can distinguish malignant from benign masses. Furthermore, the relative nature of radiomic features between hypo- and hyperechogenic components was also informative, as indicated by using the ratio of effective diameter, areas, difference entropy, and the correlation between the components. The two most important features, from Shapley analysis, were the ratio of the difference entropy between the two echogenic components and the proportion of the effective diameter of them. Overall, these features showed that the edges and the relative nature of intra-mass components enable adnexal mass evaluations that reflect the internal mass architecture. We will continue to investigate these features in future studies.

There were some limitations to our study. First, there were a limited number of masses available for analysis. This is because we incorporated strict clinical and technical inclusion criteria for the masses and images used in the segmentation and classification pipeline. We did this to reduce confounding factors such as variation in image acquisition parameters, a process that is important in AI/CADx development. Future studies will include investigation into out-of-distribution performance, such as on masses with undefined borders on imaging and pelvic pathologies that do not arise from the adnexal region (e.g., pedunculated fibroids and appendiceal tumors). Second, the size of the dataset was also influenced by both the retrospective single-center design and the accrual of cases at our institution. We will continue to evaluate our pipeline as additional cases accrue at our institution and ultimately initiate a prospective multi-center study using diverse image databases. Third, the AI/CADx model used a binary feature for the presence of solid components of the masses, derived from expert review of the case. In the future, the discovery and evaluation of solid components could be automated.

**Conclusions**

A hybrid AI/CADx pipeline incorporating automatic external mass border segmentation, automatic physics-driven internal echogenic component segmentation, and radiomic feature analysis specific to the components and their relative nature distinguishes between malignant and benign masses with very high sensitivity and relatively high specificity. This hybrid AI/CADx pipeline could potentially serve as a second reader to ensure no malignant tumor will be missed. It may also reduce user variability and reflect the mass's heterogeneous architecture. These results highlight the importance of component-based analysis of adnexal masses on ultrasound imaging for automatic assessments. Our study results support further evaluation of the hybrid pipeline on expanded cohorts (including patients managed conservatively), variations in image acquisition, and validation using independent datasets. We will also continue to investigate additional approaches, such as deep learning for classification or end-to-end methodologies.

**Acknowledgments:** The authors thank Gail Isenberg for editing the manuscript. We also thank Tricia L. Chartrand, Chantelle D. Burns, and Kelly K. DeSantiago from the University of Chicago Department of Obstetrics and Gynecology for their assistance in dataset collection. We also thank Nick Gruszauskas from the University of Chicago Human Imaging Research Office at the University of Chicago for additional assistance in dataset collection.

H.W. is supported by the Roswell Park-University of Chicago Ovarian Cancer SPORE Career Enhancement Program (NCI Grant No. 2P50CA159981), the NCI Center Support Grant to the University of Chicago Medicine Comprehensive Cancer Center (Grant No. 3P30CA014599), and the Minnesota Ovarian Cancer Alliance/Any Mountain. H.W., H.L., and M.L.G. are supported by the Medical Imaging and Data Resource Center, which is funded by the National Institute of Biomedical Imaging and Bioengineering (Contract No. 75N92020D00021), ARPA-H, and the Department of Radiology, University of Chicago. E.L. was supported by The Honorable Tina Brozman Foundation for Ovarian Cancer Research, Tina's Wish, the Chicago Lying-In 75th Anniversary fund, and NIH/NCI Grant No. R35CA264619.

**Conflicts of Interest:** J.S.A. receives royalties from UpToDate that are unrelated to this work. L.L. and H.L. receive royalties through the University of Chicago Polsky Center for Entrepreneurship and Innovation. M.L.G. is a stockholder in R2 technology/Hologic and shareholder in QView, receives royalties from various companies through the University of Chicago Polsky Center for Entrepreneurship and Innovation, and was a cofounder in Quantitative Insights (now Qlarity Imaging). E.L. receives research funding to study the biology of ovarian cancer from AbbVie through the University of Chicago that is unrelated to this work. Some of the information in this manuscript is the subject of a patent application filed and owned by The University of Chicago. It is in the University of Chicago Conflict of Interest Policy that investigators disclose publicly actual or potential significant financial interest that would reasonably appear to be directly and significantly affected by the research activities. No other disclosures were reported.

**Figures**

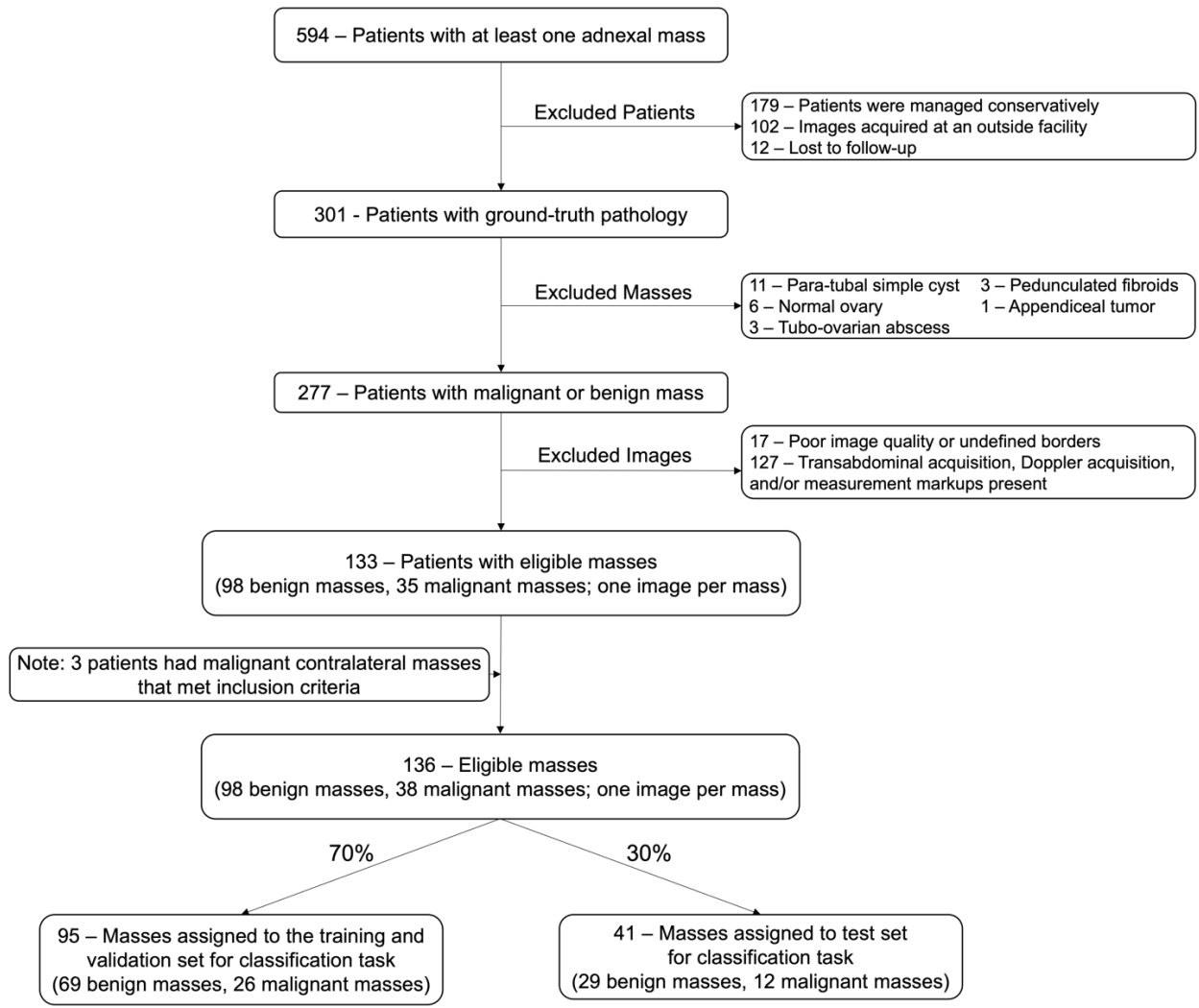

**Figure 1:** Flowchart showing exclusion criteria and resulting eligible cases and masses.

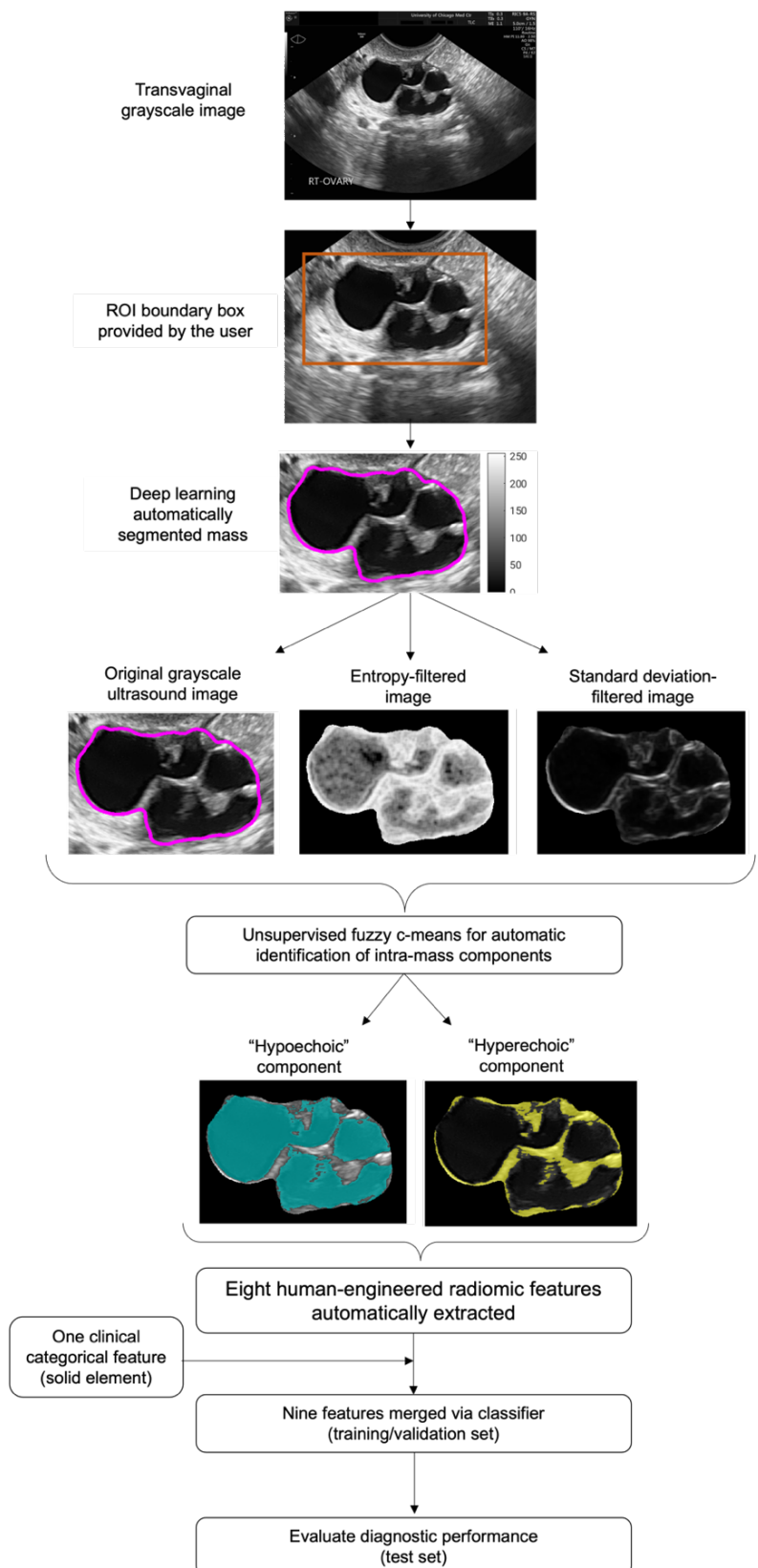

**Figure 2:** AI/CADx pipeline for adnexal mass diagnosis. AI/CADx: Artificial intelligence/computer-aided diagnosis.

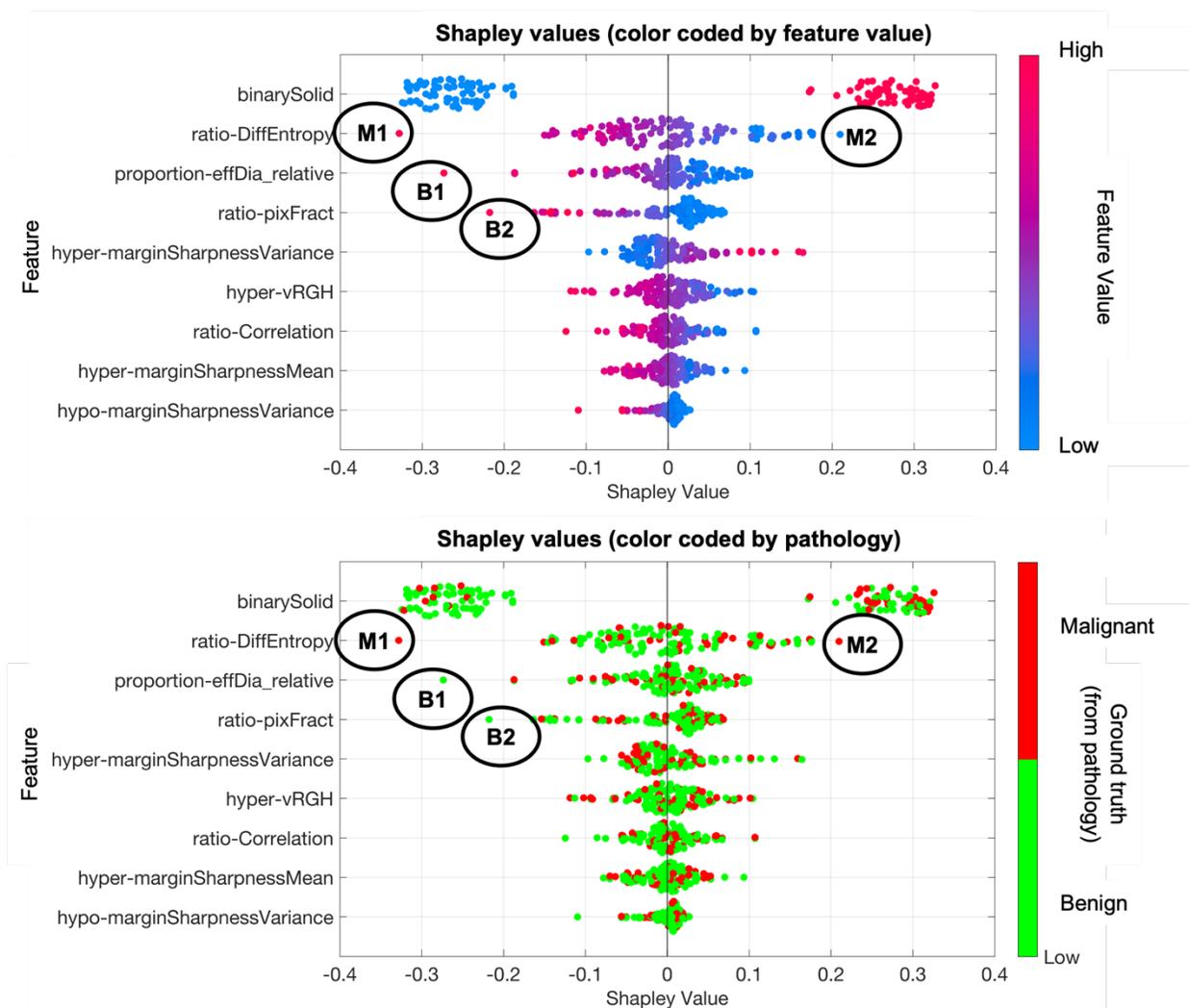

**Figure 3:** Shapley bee swarm plots for the features used in the AI/CADx classification of adnexal masses as malignant or benign on ultrasound images. The results are shown (top) color coded by feature value (the traditional presentation of Shapley values) and (bottom) color coded by pathology. Note that in Shapley bee swarm plots, each row is comprised of data points for all of the masses, and the figure format does not track each mass across the rows. The color coding by pathology emphasizes that for this classification task on this modality, the importance of features varies among benign and malignant masses, highlighting the utility of the echogenic component-based approach for heterogeneous adnexal masses. Four example masses (two malignant masses labeled M1 and M2 and two benign masses labeled B1 and B2) are identified for further examination in Figure 4.

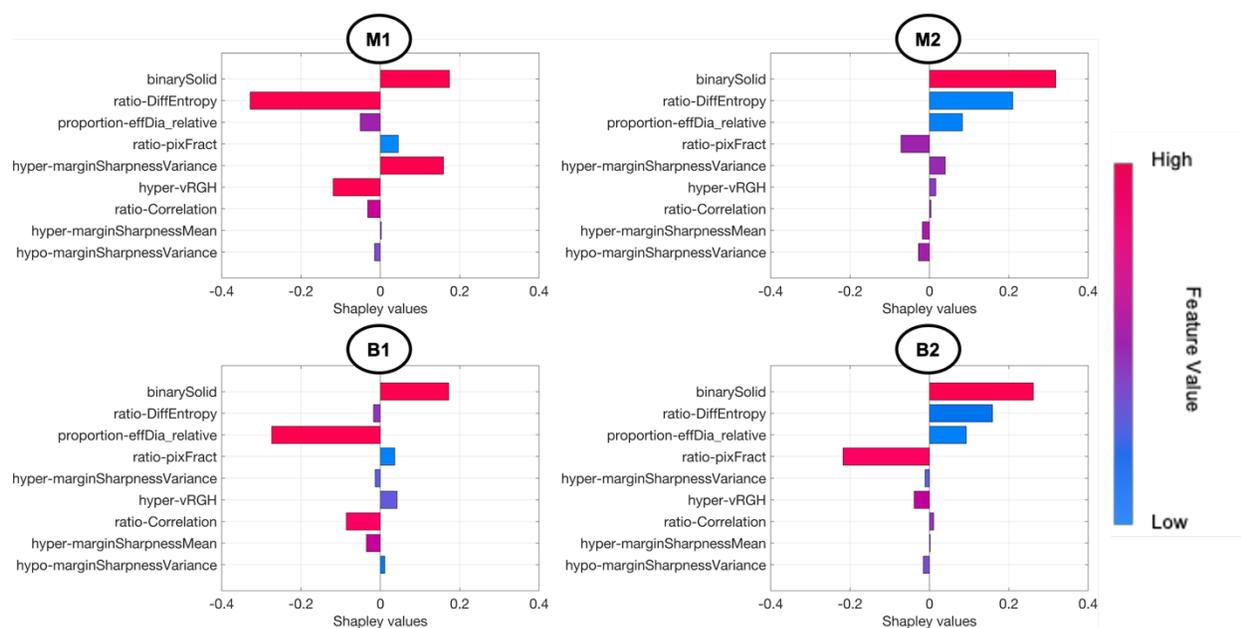

**Figure 4:** Shapley feature importance bar charts for two example malignant (M1 and M2) and two example benign (B1 and B2) masses. The mass labels are the same as in Figure 3. These results demonstrate how the importance of the features for classification varies both between and within mass types.

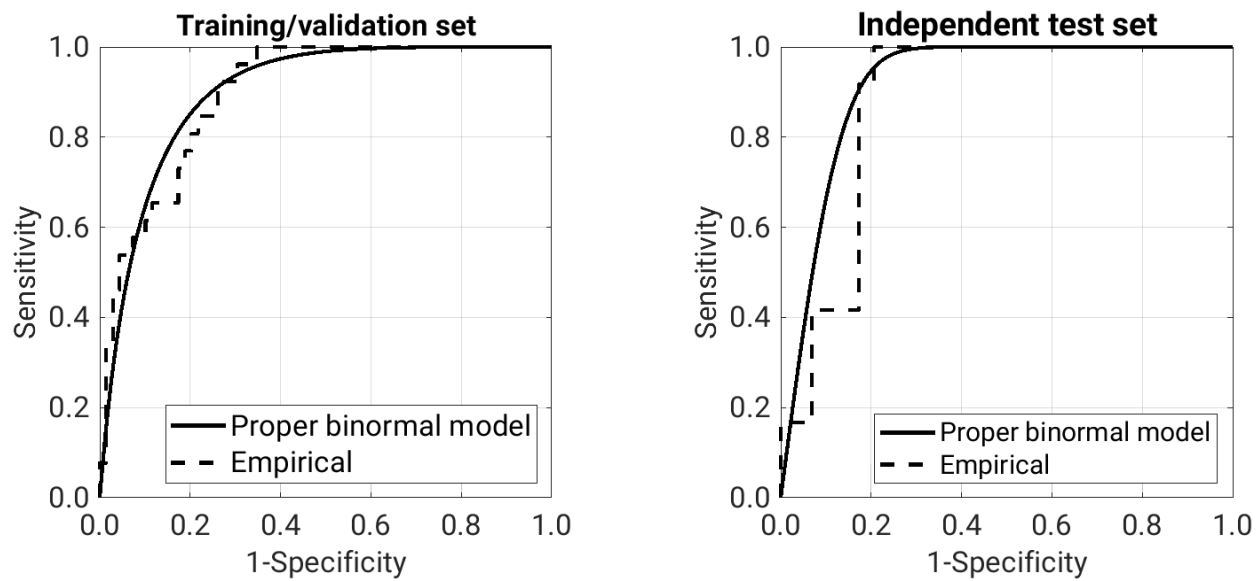

**Figure 5:** ROC analysis in the task of classifying adnexal masses as malignant or benign. Both the proper binormal model and empirical curves are shown. The AUC for the proper binormal model was (median, [95% CI]) 0.90 [0.84, 0.95] in the training/validation set and 0.93 [0.83, 0.98] in the independent test set. ROC: receiver operating characteristic. AUC: area under the receiver operating characteristic curve

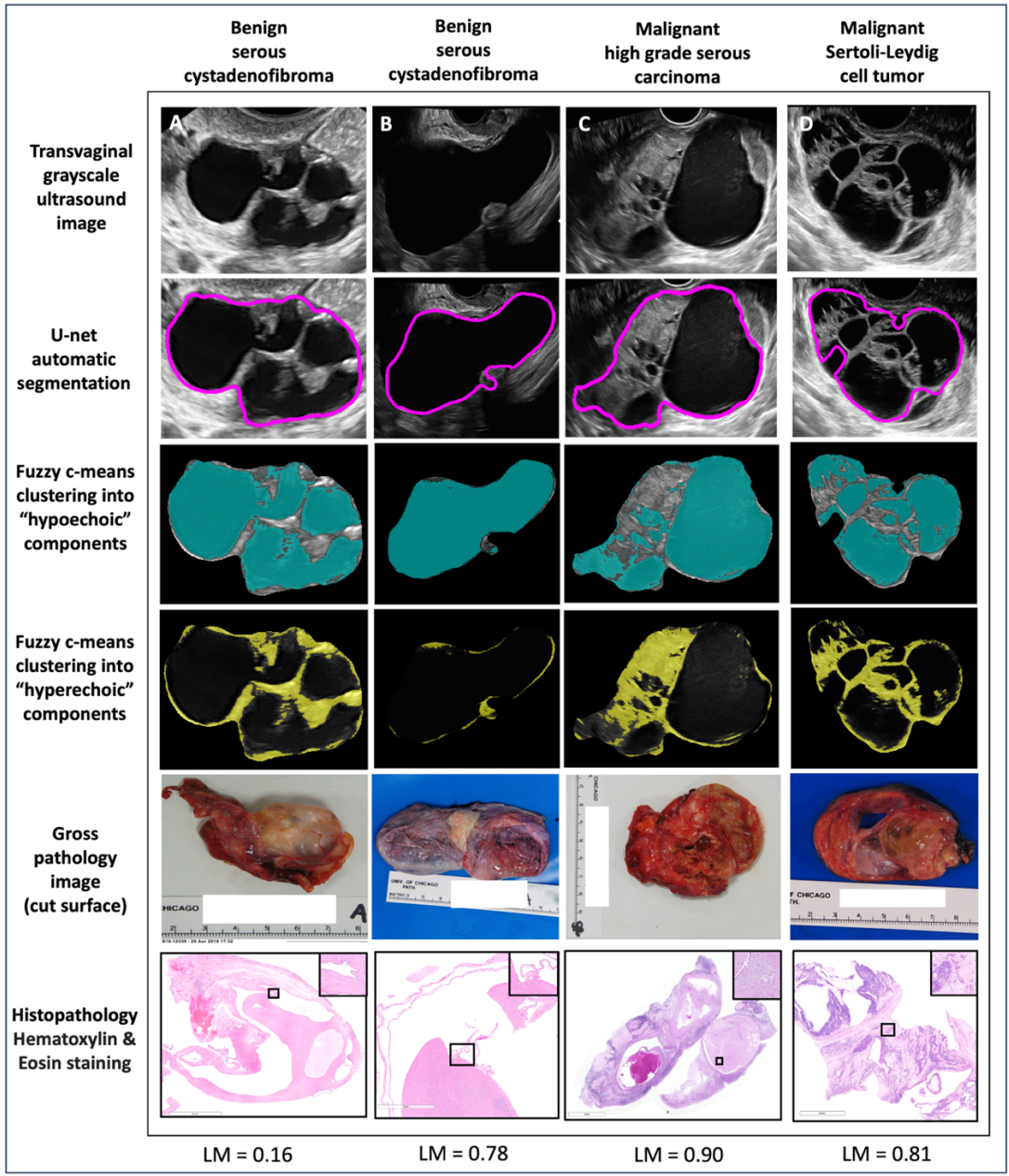

**Figure 6:** Sonographic, AI/CADx-based automatic segmentation, component-based clustering, and histopathology examples of individual masses in the test set. Images of two benign (A,B) and two malignant (C,D) ovarian masses and their corresponding likelihood of malignancy (LM) from prediction as malignant or benign by the AI/CADx model are shown. AI/CADx: artificial intelligence/computer-aided diagnosis

Note. – Pathology case numbers are obscured from the images according to HIPAA regulations.

**Supplemental materials**

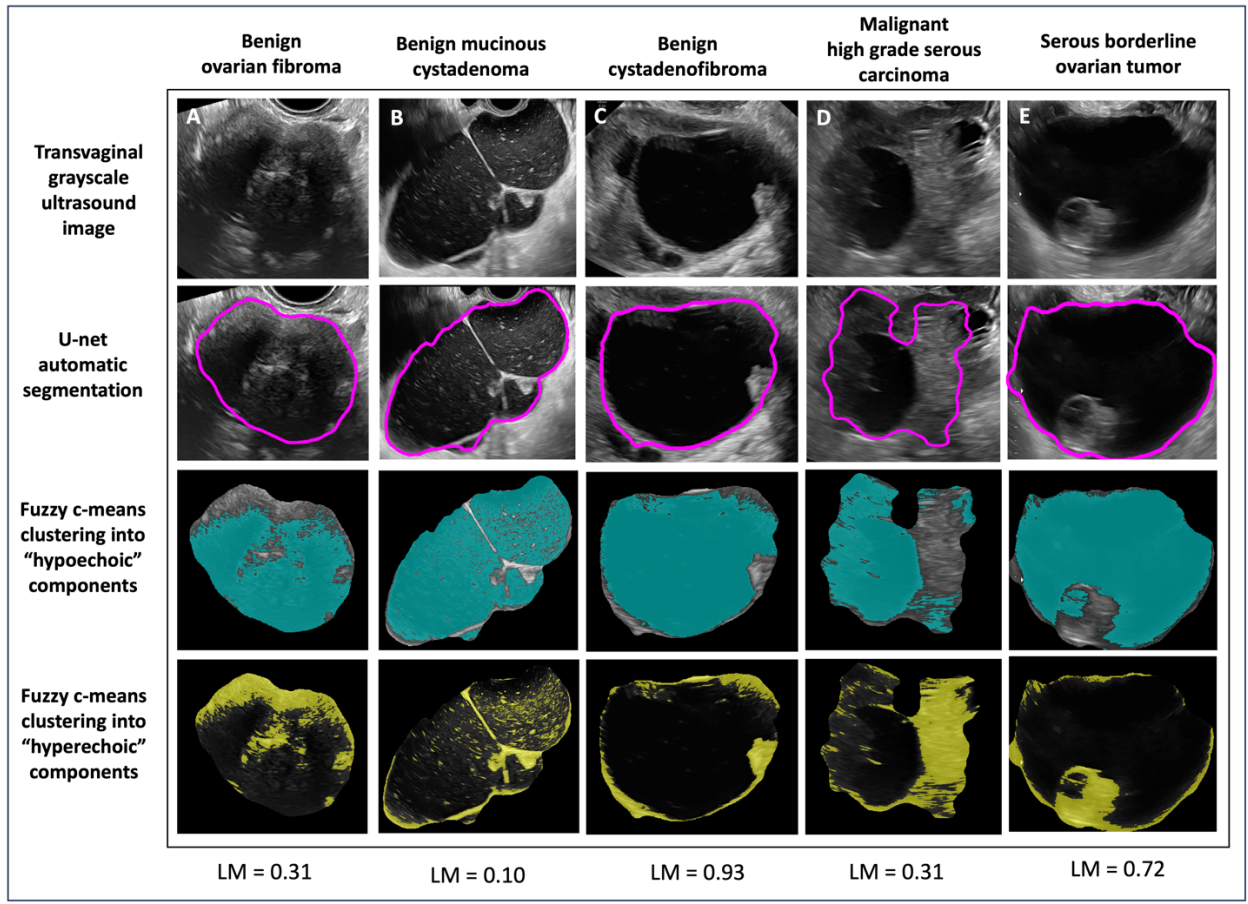

**Supplemental materials Figure 1:** Sonographic, AI/CADx-based automatic segmentation and component-based clustering of individual masses in the training/validation set. Images of three benign (A,B,C) and two malignant/borderline (D,E) ovarian masses and their corresponding likelihood of malignancy (LM) from prediction as malignant or benign by the AI/CADx model are shown. AI/CADx: artificial intelligence/computer-aided diagnosis.

**Tables**

**Table 1:** Demographics and clinicopathological characteristics of patients by training/validation and test sets (N=133 patients).

| Parameter | Training/validation set | Test set |
|---|---|---|
| Number of patients | 92 | 41 |
| Mean age (years) [a] | 46.7 (16.3) | 46.6 (14.8) |
| Age range (years) | 20-82 | 20-79 |
| Menopausal status | | |
|    Premenopausal | 57 (62.0) | 23 (56.1) |
|    Postmenopausal | 35 (38.0) | 18 (43.9) |
| Race | | |
|    Black patients | 39 (42.4) | 18 (43.9) |
|    White patients | 37 (40.2) | 14 (34.1) |
|    Other patients | 13 [b] (14.1) | 4 [c] (9.8) |
|    Declined | 3 (3.3) | 5 (12.2) |
| Ethnicity | | |
|    Hispanic or Latino | 2 (2.2) | 4 (9.8) |
|    Not Hispanic or Latino | 86 (93.5) | 32 (78.0) |
|    Declined | 4 (4.3) | 5 (12.2) |
| CA-125 Tumor Marker | | |
|    Available (yes) | 59 (64.1) | 24 (58.5) |
|    Elevated (>35 U/mL) | 21 (22.8) | 10 (24.3) |
|    Mean[a] | 254.6 (804.4) | 188.04 (429.9) |

Note. – Unless otherwise indicated, data are numbers of patients and data in parentheses are percentages. Percentages may not add up to 100% due to rounding. The dataset was split into training/validation and test sets by patient. Three patients had bilateral malignant masses.

[a] Data in parentheses are standard deviation
[b] Other race groups include Asian/ Mideast Indian (n=9), Native Hawaiian or other Pacific Islander (n=1), and more than 1 race (n=3).
[c] Other race groups include Asian/ Mideast Indian (n=2) and more than 1 race (n=2) patients.

**Table 2:** Clinicopathological characteristics of masses training/validation and test sets (N=136 masses)

| Parameter | Training/validation set | Test set |
|---|---|---|
| Number of masses | 95 | 41 |
| Mass type | | |
|   Benign | 69 (72.6) | 29 (70.7) |
|   Borderline | 4 (4.2) | 2 (4.9) |
|   Malignant | 22 (23.2) | 10 (24.4) |
| Benign pathology subtype | | |
|   Endometrioma | 16 (16.8) | 5 (12.2) |
|   Mature Teratoma | 13 (13.7) | 6 (14.6) |
|   Cystadenofibroma | 9 (9.5) | 4 (9.8) |
|   Serous cystadenoma | 9 (9.5) | 4 (9.8) |
|   Other benign ovarian pathology | 6 (6.3) | 3 (7.3) |
|   Hemorrhagic cyst, hemorrhagic corpus luteum | 5 (5.3) | 2 (4.9) |
|   Fibroma, thecoma, fibrothecoma | 3 (3.2) | 1 (2.4) |
|   Mucinous cystadenoma | 3 (3.2) | 2 (4.9) |
|   Hydrosalpinx | 2 (2.1) | 1 (2.4) |
|   Other benign pathology | 2 (2.1) | 0 (0.0) |
|   Struma ovarii | 1 (1.1) | 0 (0.0) |
|   Seromucinous cystadenoma | 0 (0.0) | 1 (2.4) |
| Malignant pathology subtype | | |
|   Serous borderline | 3 (3.2) | 1 (2.4) |
|   Mucinous borderline | 1 (1.1) | 1 (2.4) |
|   High grade serous carcinoma | 9 (9.5) | 3 (7.3) |
|   Low grade serous carcinoma | 1 (1.1) | 1 (2.4) |
|   Clear cell carcinoma | 2 (2.1) | 1 (2.4) |
|   Endometrioid carcinoma | 1 (1.1) | 1 (2.4) |
|   Mucinous carcinoma | 1 (1.1) | 1 (2.4) |
|   Granulosa cell tumor | 2 (2.1) | 1 (2.4) |
|   Immature teratoma | 1 (1.1) | 0 (0.0) |
|   Sertoli-Leydig cell tumor | 1 (1.1) | 1 (2.4) |
|   Sarcoma | 1 (1.1) | 0 (0.0) |
|   Metastases to the ovary | 3 (3.2) | 1 (2.4) |

Note. – Unless otherwise indicated, data are numbers of masses, and data in parentheses are percentages. Percentages may not add up to 100% due to rounding. The dataset was split into training/validation and test sets by patient. Three patients had bilateral malignant masses.

**Table 3:** Descriptions of radiomic features used in the study.

| Feature name | Feature source | Category |
|---|---|---|
| Margin sharpness variance<br>*(Variance of the image gradient at the component margin)* | Hypoechoic component | Morphology |
| Margin sharpness variance<br>*(Variance of the image gradient at the component margin)* | Hyperechoic component | Morphology |
| Margin sharpness mean<br>*(Mean of the image gradient at the component margin)* | Hyperechoic component | Morphology |
| Variance of the radial gradient histogram | Hyperechoic component | Morphology |
| Effective diameter<br>*(Diameter of a circle with the same area as the component)* | Proportion between components | Shape |
| Pixel fraction<br>*(ratio of the number of pixels in the component to all pixels within the tumor)* | Hypoechoic component | Size |
| Difference entropy<br>*(Measure of the randomness of the difference of neighboring pixels' gray levels)* | Ratio between components | Texture |
| Correlation<br>*(Measure of the joint probability occurrence of the specified pixel pairs)* | Ratio between components | Texture |

**Table 4:** Median diagnostic performance at target 95% sensitivity (determined in the training/validation set) of the AI/CADx model for classifying benign and malignant adnexal masses

| Set | Sensitivity (%) | Specificity (%) | Positive predictive value (%) | Negative predictive value (%) | Accuracy (%) |
|---|---|---|---|---|---|
| Training/validation | 95.0 (25/26) [94.7,95.3] | 70 (48/69) [51, 83] | 54 (25/46) [42,68] | 97 (48/49) [96, 98] | 76 (73/95) [63, 87] |
| Test | 99.6 (12/12) [87.8,100] | 71 (23/29) [53, 84] | 58 (12/18) [47, 71] | 99.7 (23/23) [93.5,100] | 79 (35/41) [67, 88] |

Note. – Numbers in parentheses are raw data from the empirical ROC curves, and numbers in brackets are 95% confidence intervals. Values are reported with three significant figures when necessary to differentiate median from confidence interval.